\documentclass[12pt, english,draftclsnofoot, onecolumn]{IEEEtran}
\usepackage{amsfonts}
\usepackage{amssymb}
\usepackage{amsmath}
\usepackage{amsthm}
\usepackage{stackrel}
\usepackage{relsize}
\usepackage{float}
\usepackage{bm}
\usepackage{babel}
\usepackage{lipsum}
\usepackage{courier}
\usepackage{lineno}
\usepackage{mathtools}
\usepackage{xcolor}
\theoremstyle{remark}
\newtheorem{rem}{\protect\remarkname}
\providecommand{\remarkname}{Remark}

\begin{document}
\title{A Comprehensive Performance Evaluation of a DF-Based Multi-Hop System Over $\alpha-\kappa-\mu$ and $\alpha-\kappa-\mu$-Extreme Fading Channels}

\author{Tau Raphael Rasethuntsa, Sandeep Kumar$^\dagger$, Manpreet Kaur$^\star$

\thanks{Tau Raphael Rasethuntsa is based in, Lenasia Extension 9, Gauteng Province, Johannesburg, South Africa and Ha Abia/Mahlabatheng, Moreneng, Maseru, Lesotho, e-mail: tauzand@gmail.com}

\thanks{Sandeep Kumar and Manpreet Kaur are with the Central Research Laboratory, Bharat Electronics Limited, Ghaziadbad, Uttar Pradesh, India, e-mail: sann.kaushik@gmail.com$^\dagger$, manpreettiet@gmail.com$^\star$}

\thanks{}}

\markboth{}
{}

\maketitle

\begin{abstract}
In this work, an integrated performance evaluation of a decode-and-forward (DF) multi-hop wireless communication system is undertaken over the non-linear generalized $\alpha-\kappa-\mu$ and $\alpha-\kappa-\mu$-Extreme fading models. Analytical formulas for the probability density function (PDF) and the cumulative distribution function (CDF) of the received signal-to-noise ratio (SNR) as well as its generalized moments and moment generating function (MGF) are derived. Based on the derived PDFs, novel closed-form expressions for traditional performance metrics such as amount of fading (AF), outage probability (OP), bit error rate (BER) under coherent and non-coherent modulation schemes as well as channel capacity under various adaptive transmission techniques are derived. Additionally, asymptotic analyses of BER based on Poincare series expansions of SNR PDFs are carried out and results show good approximations for low SNR regimes. The correctness of the proposed solutions has been corroborated by comparing them with Monte Carlo simulation results.
\end{abstract}

\begin{IEEEkeywords}
Multi-hop system, decode-and-forward relay, bivariate Fox H-function, non-linear generalized fading, Marcum Q-function, Nuttall Q-function
\end{IEEEkeywords}

\IEEEpeerreviewmaketitle

\section{Introduction}
The use of multi-hop communication systems as a way of increasing network coverage area and energy-efficiency with high data rates is gaining importance in next-generation wireless networks. In a multi-hop wireless communication system, information is transmitted from a source device to a destination device through several intermediate nodes, referred to as relays. Based on the nature and complexity of the relaying system, multi-hop communication systems can be classified into two categories, namely (i) decode-and-forward (DF) and (ii) amplify-and-forward \cite{kramer2005cooperative}. For a DF-based multi-hop system, the relaying node receives the encoded signal, decodes it to regenerate the original symbol and forwards the newly encoded signal to the next node \cite{islam2013error}. DF-based multi-hop systems have several applications in ad-hoc networks, sensor networks and microwave links among many others \cite{cao2012performance}.

The performance analysis of multi-hop communication systems over numerous fading distribut- ions has been well documented in the literature \cite{cao2012performance,farhadi2008ergodic,bhatnagar2013capacity,dixit2014performance,tha2017performance}. The ergodic capacity of a multi-hop relaying system over Rayleigh fading channels has been studied in \cite{farhadi2008ergodic}. The end-to-end performance of a DF relaying system based on Ricean and Nakagami-$m$ fading channels was explored in \cite{bhatnagar2013capacity}. Yang et al. \cite{yang2005average} studied the unified performance analysis of a DF-based multi-hop system based on second order statistics in average outage duration (AOD) and level crossing rate (LCR) over Nakagami, Rayleigh and Rician fading channels. The performance study of a multi-hop communication system with regenerative relays has been investigated over generalized $\eta-\mu,$ $\kappa-\mu$ and $\alpha-\mu$ fading channels in \cite{dixit2014performance}, \cite{tha2017performance} and \cite{dixit2017exact}, respectively. Using traditional performance metrics as well as LCR and amount of fade duration (AFD), Cao et al. \cite{cao2012performance} presented the performance analysis of end-to-end wireless links over the Generalized-$K$ fading channel.

The modeling of mobile radio channel is a challenging task as the wireless signal is subjected to various hurdles/corruptions in its propagation path from
the transmitter to the receiver \cite{porto2016phase}. Accurate characterization of the wireless channel is therefore of paramount importance for the realistic assessment of wireless systems performances. Several statistical distributions were developed in the literature to model small-scale fluctuations of the fading channel envelope \cite{kumar2018performance}. For instance, motivated by the pioneering works in \cite{yacoub2007kappa} and \cite{rabelo2009mu}, the non-linear generalized $\alpha-\kappa-\mu$ and $\alpha-\kappa-\mu$-Extreme fading models were proposed in \cite{fraidenraich2006alpha},  \cite{sofotasios2011alpha} and \cite{Rasethuntsa2019alpha}. The $\alpha-\kappa-\mu$ distribution is a very flexible model that does not assume homogeneous scattering and can account well for non-linearity of multi-path fading \cite{KUMAR201826}. The model includes as special cases several other well-known distributions. In particular, when $\alpha=2$, the $\alpha-\kappa-\mu$ distribution reduces to the $\kappa-\mu$ distribution \cite{yacoub2007kappa} and for $\kappa\rightarrow0,$ the $\alpha-\mu$ is obtained \cite{tha2017performance}.   The $\alpha-\kappa-\mu$-Extreme fading model is derived from the $\alpha-\kappa-\mu$ distribution by allowing $\mu\rightarrow0$ and $\kappa\rightarrow\infty$ in order to better characterize small-scale variations of mobile radio propagation under non-linear severe multi-path fading effects usually encountered in enclosed environments \cite{sofotasios2011alpha}. For $\alpha=2,$ the $\alpha-\kappa-\mu$-Extreme distribution reduces to a linear severe fading model, the $\kappa-\mu$-Extreme distribution introduced by Rabelo et al. \cite{rabelo2009mu}.
Although there are many works related to the study of DF-based multi-hop relaying systems over different fading channels, none of the previous works have investigated the end-to-end performance over $\alpha-\kappa-\mu$ and $\alpha-\kappa-\mu$-Extreme fading channels. Motivated by this void, we have derived the closed-form mathematical expressions for amount of fading (AF), outage probability (OP), bit error rate (BER) (coherent and non-coherent) and channel capacity for ene-to-end DF relaying system under various adaptive transmissions over $\alpha-\kappa-\mu$ and $\alpha-\kappa-\mu$-Extreme fading channels. The expressions derived here are presented in a simple manner to optimize clarity and readability. Furthermore, the results produced here are generalized expressions and valid for other fading distributions as special cases.
\flushbottom
The rest of the paper is organized as follows : The system and channel model are described in Section \ref{sec:1}. In Section \ref{sec:2}, we provide the closed-form expressions for the probability density function (PDF), cumulative distribution function (CDF) generalized moments and moment generating function of the end-to-end signal-to-noise ratio (SNR). Different performance metrics for the DF-based multi-hop wireless system over $\alpha-\kappa-\mu$ and $\alpha-\kappa-\mu$-Extreme fading channels are presented in Section \ref{sec:3}. Asymptotic analysis of BER based on Poincare expansions of fading model  (PDFs) is carried out in Section \ref{sec:4}. Performance evaluation, simulation results and discussions are provided in Section \ref{sec:5}. Finally, we round up the paper in Section \ref{sec:6} with conclusions and possible future considerations.

\section{Model Formulation and Description}\label{sec:1}

\subsection{System and Channel Models}
\begin{figure}[!htb]
\vspace{-0.8cm}
\makebox[\columnwidth]{\includegraphics[clip,width=21pc]{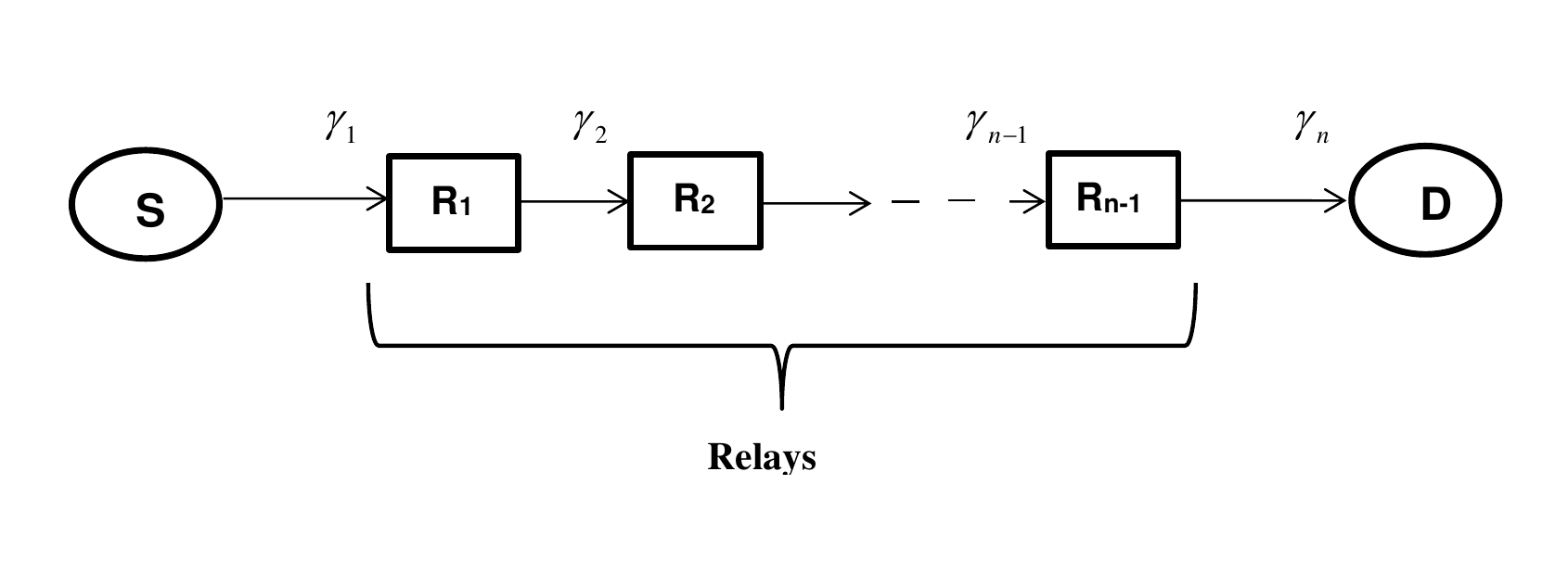}}
\vspace{-1cm}
\caption{A $n$-link DF-based multi-hop system.}\label{fig:0}
\end{figure} Consider an $n$-hop system shown in Figure \ref{fig:0} transmitting information from the source $S$ to the destination $D$ via $n-1$ intermediate relays $R_{1},R_{2},\ldots,R_{n-1}$ in time-sharing principles.  If the SNR of the $i$-th hop is denoted by $\gamma_{i},$ with PDF $f_{\gamma_{i}},$ $i=1,\ldots n,$ the statistical properties of the end-to-end system SNR are then determined by the SNR of the last hop $\gamma_{n}.$ According to \cite{beaulieu2006closed}, the PDF of the end-to-end system SNR $\gamma$ can be written as\vspace{1.5mm}\begin{equation}\label{definition-pdf-of-system-snr}
  f_{\gamma}(\gamma)=\mathcal{A}\delta(\gamma)+\bar{\mathcal{A}}f_{\gamma_{n}}(\gamma),
\end{equation}\vspace{1.5mm}where $\mathcal{A}$ denotes the probability of the first outage occurring for the first $n-1$ hops, $\bar{\mathcal{A}}=1-\mathcal{A}$ and $\delta(\cdot)$ is the Dirac delta function. The probability $\mathcal{A}$ is given by \cite{beaulieu2006closed}\vspace{1.5mm}\begin{equation}\label{A-outage-probability}
\begin{aligned}
  \mathcal{A}&=1-Pr\left(\gamma_{1}>\gamma_{th},\gamma_{2}>\gamma_{th},\cdots,\gamma_{m-1}>\gamma_{th}\right)\\
  &=1-\smashoperator[r]{\prod_{k=1}^{m-1}}Pr\left(\gamma_{n}>\gamma_{th}\right)=1-\smashoperator[r]{\prod_{k=1}^{m-1}}\left[ 1-F_{\gamma_{n}}(\gamma_{th})\right],
  \end{aligned}
\end{equation}\vspace{1.5mm}where $F_{\gamma_{n}}$ is the CDF of $\gamma_{n}.$

\subsubsection{The $\alpha-\kappa-\mu$ and $\alpha-\kappa-\mu$-Extreme fading models}
The PDF of the instantaneous SNR of the $n$-th hop $\gamma_{n}>0$ following the $\alpha_{n}-\kappa_{n}-\mu_{n}$ distribution is given by\vspace{1.5mm}\begin{equation}\label{pdf-of-aku}
  f_{\gamma_{n}}(\gamma_{n})=\mathcal{K}_{n}\gamma_{n}^{(\frac{\alpha_{n}}{2})\omega_{n}-1}e^{-\sigma_{n}\gamma_{n}^{\frac{\alpha_{n}}{2}}}I_{\mu_{n}-1}\left(\sqrt{\theta_{n}\gamma_{n}^{\frac{\alpha_{n}}{2}}}\right),
\end{equation}\vspace{1.5mm}where, in order to improve clarity in presentation, we have set \vspace{1.5mm}\begin{equation}\label{reductedsymbols}
\begin{aligned}
  \mathcal{K}_{n}=\dfrac{\alpha_{n}\sigma_{n}^{\omega_{n}}e^{-\mu_{n}\kappa_{n}}}{2(\mu_{n}\kappa_{n})^{\omega_{n}-1}},\;\omega_{n}&=\dfrac{\mu_{n}+1}{2},\; \sigma_{n}=\dfrac{\mu_{n}(\kappa_{n}+1)}{\bar{\gamma}_{n}^{\frac{\alpha_{n}}{2}}},\\
  \;\theta_{n}&=4\mu_{n}\kappa_{n}\sigma_{n}
  \end{aligned}
\end{equation}\vspace{1.5mm}with $\alpha_{n},\mu_{n},\kappa_{n}>0.$ The average SNR per symbol $\bar{\gamma}_{n}$ is given by\vspace{1.5mm}\begin{equation*}
  \bar{\gamma}_{n}=\dfrac{e^{\mu_{n}\kappa_{n}}\left[\mu_{n}(\kappa_{n}+1)\right]^{\frac{2}{\alpha_{n}}}}{\Gamma\left( \frac{2}{\alpha_{n}}+\mu_{n}\right)\tilde{\Phi}\left(\frac{2}{\alpha_{n}}+\mu_{n};\mu_{n};\mu_{n}\kappa_{n} \right)}\dfrac{E_{b}}{N_{0}},
\end{equation*}\vspace{1.5mm}where $E_{b}$ denotes the energy per symbol and $N_{0}$ is the single-sided power spectral density. Using the method of \cite{kumbhani2017mimo}, one can easily show that the CDF of $\gamma_{n}$ is written as\vspace{1.5mm}\begin{equation}\label{cdf-of-aku}
  F_{\gamma_{n}}(\gamma_{n})=1-Q_{\mu_{n}}\left(\sqrt{2\kappa_{n}\mu_{n}},\sqrt{2\sigma_{n}\gamma_{n}^{\frac{\alpha_{n}}{2}}} \right),\;\gamma_{n}>0,
\end{equation}\vspace{1.5mm}where $Q(\cdot,\cdot)$ is the generalized Marcum Q-function.
The PDF of the instantaneous SNR of the $n$-th hop $\gamma_{n}>0$ following the $\alpha_{n}-\kappa_{n}-\mu_{n}$-Extreme distribution is given by\vspace{1.5mm}\begin{equation}\label{PDFExtreme}
  g_{\gamma_{n}}(\gamma)=\dfrac{a_{n}\gamma_{n}^{\alpha_{n}/4-1}}{e^{b_{n}\gamma_{n}^{\frac{\alpha_{n}}{2}}}}I_{1}\left(\sqrt{c_{n}\gamma_{n}^{\frac{\alpha_{n}}{2}}} \right)+\dfrac{e^{-2m_{n}}}{2\sqrt{\gamma_{n}\bar{\gamma}_{n}}}\delta\left(\sqrt{\dfrac{\gamma_{n}}{\bar{\gamma}}}\right),
\end{equation}\vspace{1.5mm}where\vspace{1.5mm}\begin{equation}\label{reductedsymbolsII}a_{n}=\dfrac{\alpha_{n} m_{n} e^{-2m_{n}}}{\bar{\gamma}_{n}^{\alpha_{n}/4}}, b=\dfrac{2m_{n}}{\bar{\gamma}_{n}^{\frac{\alpha_{n}}{2}}}, c_{n}=\dfrac{(16m_{n}^2)}{\bar{\gamma}_{n}^{\frac{\alpha_{n}}{2}}},\end{equation}\vspace{1.5mm}with $\alpha_{n},\mu_{n},\kappa_{n}>0$ and $m_{n}=\mu_{n}(\kappa_{n}+1)^2/(2\kappa_{n}+1)$ as the Nakagami parameter which is inversely proportional to the fading severity. The average SNR per symbol is given by\vspace{1.5mm}\begin{equation*}
    \bar{\gamma}_{n}=\dfrac{e^{2m_{n}}(2m_{n})^{\frac{2}{\alpha_{n}}-1}}{\Gamma\left(\frac{2}{\alpha
    _{n}}+\mu_{n}\right)\Phi\left(\frac{2}{\alpha_{n}}+\mu_{n};\mu_{n};\mu_{n}\kappa_{n} \right)}\dfrac{E_{b}}{N_{0}}.
\end{equation*}\vspace{1.5mm}The CDF of the $\alpha-\kappa-\mu$-Extreme fading model can also be derived using the method of \cite{kumbhani2017mimo} resulting in\vspace{1.5mm}\begin{equation}\label{CDFXtreme}
   G_{\gamma}(\gamma)=1-Q_{0}\left(2\sqrt{m_{n}},2\sqrt{m_{n}(\gamma/\bar{\gamma})^{\frac{\alpha}{2}}} \right).
 \end{equation}\vspace{1.5mm}The symbols in \eqref{reductedsymbols} and \eqref{reductedsymbolsII} will be employed throughout the paper unless the original notation produces shorter expressions.
\begin{rem}\label{remark:1} It is also worth noting that the proposed PDFs representations renders, purely on a symbolic level, the $\alpha-\kappa-\mu$-Extreme fading model PDF without the extra $\delta(\cdot)$ term as a special case of the PDF of the $\alpha-\kappa-\mu$ model in a sense that
$\mathcal{K}_{n}=a_{n},$ $\omega_{n}=1/2,$ $\sigma_{n}=b_{n},$ $\theta_{n}=c_{n}$ and $\mu_{n}=2.$ This observation will be particularly useful for reducing redundancy in the forthcoming derivations.
\end{rem}

\section{Statistical Properties}\label{sec:2}

\subsection{PDF and CDF of End-to-End SNR}
\textbf{Case I} : Assume that the SNRs of the $n$ hops are independent and not necessarily identically distributed (INID) random variables following the $\alpha-\kappa-\mu$ distribution. Then we have from \eqref{pdf-of-aku} and \eqref{cdf-of-aku} that the PDF and CDF of the end-to-end SNR $\gamma$ must be given respectively by \vspace{1.5mm}\begin{equation}\label{pdf-of-SNR-final}
  p_{\gamma}(\gamma)=\mathcal{A}\delta(\gamma)+\mathcal{K}_{n}\bar{\mathcal{A}}\dfrac{\gamma^{(\frac{\alpha_{n}}{2})\omega_{n}-1}}{e^{\sigma_{n}\gamma^{\frac{\alpha_{n}}{2}}}}I_{\mu_{n}-1}\left(\sqrt{\theta_{n}\gamma^{\frac{\alpha_{n}}{2}}}\right)
\end{equation}\vspace{1.5mm}where $\mathcal{A}=1-\prod_{k=1}^{n-1}Q_{\mu_{k}}\left(\sqrt{2\kappa_{k}\mu_{k}},\sqrt{2\sigma_{k}\gamma_{th}^{\frac{\alpha_{k}}{2}}}\right)$ and \vspace{1.5mm} \begin{equation}\label{CDF-of-system-SNR-f-final}
P_{\gamma}(\gamma)=1-\bar{\mathcal{A}}Q_{\mu_{n}}\left(\sqrt{2\kappa_{n}\mu_{n}},\sqrt{2\sigma_{n} \gamma^{\frac{\alpha_{n}}{2}}} \right).
\end{equation}\vspace{1.5mm}\textbf{Case II} : On the other hand if the SNRs of the $n$ hops are INID random variables following the $\alpha-\kappa-\mu$-Extreme distribution, then we have from \eqref{PDFExtreme} and \eqref{CDFXtreme} that the PDF of the end-to-end SNR $\gamma$ must be given by\vspace{1.5mm}\begin{equation}\label{pdf-of-SNR-final-xtreme}
  p_{\gamma}^{e}(\gamma)=\mathcal{A}^{e}\delta(\gamma)+\bar{\mathcal{A}}^e\left[\dfrac{a_{n}\gamma^{\frac{\alpha_{n}}{4}-1}}{e^{b_{n}\gamma^{\frac{\alpha_{n}}{2}}}}I_{1}\left(\sqrt{c_{n}\gamma^{\frac{\alpha_{n}}{2}}} \right)+\delta^*(\gamma)\right]
\end{equation}\vspace{1.5mm}where$\delta^*(\gamma)=\dfrac{e^{-2m_{n}}}{2\sqrt{\gamma\bar{\gamma}_{n}}}\delta\left(\sqrt{\dfrac{\gamma_{n}}{\bar{\gamma}}}\right)\\$ and $\mathcal{A}^{e}=1-\prod_{k=1}^{n-1}Q_{0}\left(2\sqrt{m_{k}},2\sqrt{m_{k}(\gamma_{k}/\bar{\gamma}_{k})^{\frac{\alpha_{k}}{2}} }\right).$ The CDF of $\gamma$ can be written as \vspace{1.5mm}\begin{equation}\label{CDF-of-system-SNR-f-final-extreme}
  P_{\gamma}^{e}(\gamma)=1-\bar{\mathcal{A}}^eQ_{0}\left(2\sqrt{m_{n}},2\sqrt{m_{n}(\gamma/\bar{\gamma})^{\frac{\alpha_{n}}{2}} }\right).
\end{equation}

\subsection{Moments of End-to-End SNR}\label{moments-I}
Generalized moments of the end-to-end SNR are useful for obtaining $r$-th-order performance metrics such as the mean, variance as well as kurtosis and skewness of the distribution of $\gamma.$ The $r$-th moment of a positive random variable $X$ with density $f_{X}$ is defined by \vspace{1.5mm}\begin{equation}\label{DefnMoment}
  \mathbb{E}\langle X^r\rangle\triangleq\int_{0}^{\infty}x^{r}f_{X}(x)dx
\end{equation}\vspace{1.5mm}where $\mathbb{E}\langle\cdot\rangle$ denotes the expectation operation. For \textbf{Case I}, substituting \eqref{pdf-of-SNR-final} into \eqref{DefnMoment} and setting $y=\gamma^{\alpha_{2}/2},$ the $r$-th moment of $\gamma$ can be written with the aid of {\cite[Eq. 6.643(2)]{gradshteyn2014table} and \cite[Eq. 6.220(2)]{gradshteyn2014table}} as\vspace{1.5mm}\begin{equation}\label{n-th-moment-of-SNR}
  \mathbb{E}\langle\gamma^r\rangle=\dfrac{\bar{\mathcal{A}}\bar{\gamma}_{n}^{r}\Gamma\left(\frac{2r}{\alpha_{n}}+\mu_{n} \right)\tilde{\Phi}\left(\frac{2r}{\alpha_{n}}+\mu_{n};\mu_{n};\mu_{n}\kappa_{n}\right)}{e^{\mu_{n}\kappa_{n}}\left[ \mu_{n}(\kappa_{n}+1)\right]^{\frac{2r}{\alpha_{n}}}}
\end{equation}\vspace{1.5mm}where $\tilde{\Phi}$ is the regularized confluent hypergeometric function {\cite[Eq. 9.210(1)]{gradshteyn2014table}}. { 
 Similarly, for \textbf{Case II}, we have the $r$-th moment of $\gamma$ as\vspace{1.5mm}\begin{equation}\label{n-th-moment-of-SNR-extreme}
  \mathbb{E}^e\langle\gamma^r\rangle=\dfrac{\bar{\mathcal{A}}^e\bar{\gamma}_{n}^{r}\Gamma\left( \frac{2r}{\alpha_{n}}+1\right)\Phi\left(\frac{2r}{\alpha_{n}}+1;2;2m_{n}\right)}{e^{2m_{n}}(2m_{n})^{\frac{2r}{\alpha_{n}}-1}}
\end{equation}\vspace{1.5mm}It is also worth noting that as expected, upon substituting $r=1$ in \eqref{n-th-moment-of-SNR} and \eqref{n-th-moment-of-SNR-extreme}, one obtains $\bar{\gamma}=\bar{\mathcal{A}}\bar{\gamma}_{n}$ and $\bar{\gamma}=\bar{\mathcal{A}}^e\bar{\gamma}_{n}$ for each case, respectively.

\subsection{Moment Generating Function}
Another useful statistical characteristic of random variables is the moment generating function (MGF) used to obtain $n$-th order moments or BER under non-coherent modulation, see Section \ref{sec:3}. The MGF of $\gamma,$ denoted by $\mathcal{M}_{\gamma},$ is defined as\vspace{1.5mm}\begin{equation}\label{DefnMGF}
  \mathcal{M}_{\gamma}(s)\triangleq\int_{0}^{\infty}e^{-s\gamma}f_{\gamma}(\gamma)d\gamma.
\end{equation}\vspace{1.5mm}For \textbf{Case I}, we substitute \eqref{pdf-of-SNR-final} into \eqref{DefnMGF} to get the integral\vspace{1.5mm}\begin{equation}\label{MGF-int}
  \mathcal{M}_{\gamma}(s)=\mathcal{A}+\mathcal{K}_{n}\bar{\mathcal{A}}\overset{\mathcal{I}_{1}}{\overbrace{\int_{0}^{\infty}\dfrac{\gamma^{(\frac{\alpha_{n}}{2})\omega_{n}-1}}{e^{s\gamma}e^{\sigma_{n}\gamma^{\frac{\alpha_{n}}{2}}}}I_{\mu_{n}-1}\left(\sqrt{\theta_{n}\gamma^{\frac{\alpha_{n}}{2}}}\right)d\gamma}}
\end{equation}\vspace{2mm}In order to evaluate $\mathcal{I}_{1},$ we set $y=\gamma^{\alpha_{n}/2}$ and employ {\cite[Eq. 1.25]{mathai2009h} and \cite[Eq. 8.4.22(1)]{brychkov1986integrals}} as well as the relationship between the Meijer-G and the Fox-H functions as\vspace{1.5mm}\begin{equation}\label{MGF-Int-For-Asymp-Anal}
\begin{aligned}
  \mathcal{I}_{1}=&\pi\int_{0}^{\infty}y^{\omega_{n}-1}e^{-\sigma_{n}y}H_{0,1}^{1,0}\left(s^{\frac{\alpha_{n}}{2}}y\left|\begin{array}{c}
-\\
\left(0,\frac{\alpha_{n}}{2}\right)
\end{array}\right.\right)\\
&\times H_{1,3}^{1,0}\left(\frac{\theta_{n}}{4} y\left|\begin{array}{c}
\left(\frac{\mu_{n}}{2},1\right)\\
(\omega_{n}-1,1),(1-\omega_{n},1),\left(\frac{\mu_{n}}{2},1\right)
\end{array}\right.\right)dy
  \end{aligned}
\end{equation}\vspace{2mm}Now, expanding the two Fox H-functions in terms of their definition as given in {\cite[Eq. 1.1.1]{mathai1978h}} and reversing the order of integration, we have that\vspace{1.5mm}\begin{equation}
\begin{aligned}
  \mathcal{I}_{1}=&\pi\dfrac{1}{(2\pi i)^2}\int_{\mathcal{L}_{r}}\int_{\mathcal{L}_{t}}\overset{\mathcal{G}}{\overbrace{\int_{0}^{\infty}y^{\omega_{n}+r+t-1}e^{-\sigma_{n} y}dy}}\\
  &\times \dfrac{\Gamma\left[-\left(\frac{\alpha_{n}}{2}\right)r\right]\Gamma(\omega_{n}-1-t)\left( s^{\frac{\alpha_{n}}{2}}\right)^r\left(\frac{\theta_{n}}{4}\right)^t drdt}{\Gamma(\omega_{n}+t)\Gamma\left(1-\frac{\mu_{n}}{2}+t\right)\Gamma\left(\frac{\mu_{n}}{2}-t\right)}
\end{aligned}
\end{equation}\vspace{2mm}where $\mathcal{L}_{r}$ and $\mathcal{L}_{t}$ are two suitable contours in the complex $r$ and $t$ planes respectively. The integral $\mathcal{G}$ can be evaluated with the aid of {\cite[Eq. 3.326(2)]{gradshteyn2014table}} to yield\vspace{1.5mm}\begin{equation}\label{MGF-Contour}
\begin{aligned}
  \mathcal{I}_{1}=&\dfrac{\pi}{\sigma_{n}^{\omega_{n}}(2\pi i)^2}\int_{\mathcal{L}_{r}}\int_{\mathcal{L}_{t}}\dfrac{\Gamma\left[-\left(\frac{\alpha_{n}}{2}\right)r\right]\Gamma(\omega_{n}-1-t) }{\Gamma(\omega_{n}+t)}\\
  &\times\dfrac{\Gamma\left(\omega_{n}+r+t \right)\left( s^{\frac{\alpha_{n}}{2}}/\sigma_{n}\right)^r\left(\frac{\theta_{n}}{4\sigma_{n}}\right)^t drdt}{\Gamma\left(1-\frac{\mu_{n}}{2}+t\right)\Gamma\left(\frac{\mu_{n}}{2}-t\right)}
\end{aligned}
\end{equation}\vspace{1.5mm}where $\Re(\omega_{n}+r+t)>0.$ Now comparing \eqref{MGF-Contour} with the definition of the bivariate Fox H-function given in {\cite[Eq. 2.57]{mathai2009h}} yields a closed-form expression for the MGF as shown in \eqref{MGF-final} of Table \ref{table:1}. Remark \ref{remark:1} suggests that the MGF for \textbf{Case II} can be obtained from \eqref{MGF-final} by inspection resulting in the MGF for \textbf{Case II} being as shown in \eqref{Final-MGF-Extreme} of Table \ref{table:1}.

 \begin{table}\fontsize{12pt}{12pt}
\caption{MOMENT GENERATING FUNCTOINS}\label{table:1}
  \centering
 \begin{minipage}{1.00\columnwidth}
    \hrule
    \hrule
    \vspace{1.5mm}\begin{equation}\label{MGF-final}\\
  \mathcal{M}_{\gamma}(s)=\mathcal{A}+\bar{\mathcal{A}}\mathcal{K}_{n}\dfrac{\pi}{\sigma_{n}^{\omega_{n}}}H_{1,0:1,0;1,3}^{0,1:1,0;1,0}\left(\left.\begin{array}{c}
\dfrac{s^{\frac{\alpha_{n}}{2}}}{\sigma_{n}}
\\
\\ \dfrac{\theta_{n}}{4\sigma_{n}}
\end{array}\right|\begin{array}{c}
 \boldsymbol{\Delta}\left(\alpha_{n},\mu_{n},\omega_{n} \right)
\\
\\ \boldsymbol{\nabla}\left(\alpha_{n},\mu_{n},\omega_{n} \right)
\end{array}\right)\\
\end{equation}\vspace{1.5mm}\begin{equation}\label{Final-MGF-Extreme}
  \mathcal{M}_{\gamma}^{e}(s)=\mathcal{A}^{e}+\bar{\mathcal{A}}^e\dfrac{a_{n}\pi}{\sqrt{b_{n}}}H_{1,0:1,0;1,3}^{0,1:1,0;1,0}\left(\left.\begin{array}{c}
\dfrac{s^{\frac{\alpha_{n}}{2}}}{b_{n}}
\\
\\ \dfrac{c_{n}}{4b_{n}}
\end{array}\right|\begin{array}{c}
 \boldsymbol{\Delta}\left(\alpha_{n},2,\frac{1}{2} \right)
\\
\\ \boldsymbol{\nabla}\left(\alpha_{n},2,\frac{1}{2} \right)
\end{array}\right).
\end{equation}\vspace{1.5mm}\begin{equation*}
   \boldsymbol{\Delta}\equiv\left(1-\omega_{n};1,1 \right):\; - \; ;\left(\frac{\mu_{n}}{2},1\right)
\end{equation*}\vspace{1.5mm}\begin{equation*}
  \boldsymbol{\nabla}\equiv -\; :\left(0,\frac{\alpha_{n}}{2} \right) ; \; (\omega_{n}-1,1),(1-\omega_{n},1),\left(\frac{\mu_{n}}{2},1\right)
\end{equation*}\vspace{1.5mm}
\hrule
\hrule
  \end{minipage}
\end{table}
\section{Performance Analysis of Multi-Hop DF-Based Relay Systems}\label{sec:3}
\subsection{Amount of Fading}
The second-order AF is a useful performance metric that can be used to account for the severity of fading. AF is defined as the ratio of the variance to the square average SNR per symbol, $AF_{\gamma}^{(2)}\triangleq\mbox{Var}(\gamma)/\bar{\gamma}^2.$ That is,\vspace{1.5mm}\begin{equation}\label{AF}
  AF_{\gamma}^{(2)}\triangleq\dfrac{\mathbb{E}\langle \gamma^2\rangle-\mathbb{E}\langle \gamma\rangle^2}{\mathbb{E}\langle \gamma\rangle^2}.
\end{equation}\vspace{1.5mm}This can be generalized to higher-order fading as $AF_{\gamma}^{(n)}\triangleq\mathbb{E}\langle\gamma^n\rangle/(\mathbb{E}\langle\gamma\rangle^n)-1.$ Substitution of $\mathbb{E}\langle\gamma\rangle$ and $\mathbb{E}\langle\gamma^2\rangle$ from \eqref{n-th-moment-of-SNR} into \eqref{AF} yields after some basic simplifications the second-order AF of $\gamma$ under \textbf{Case I} as\vspace{1.5mm}\begin{equation}\label{second-order-AF}
  AF_{\gamma}=\dfrac{e^{\mu_{n}\kappa_{n}}\Gamma\left(\frac{4}{\alpha_{n}}+\mu_{n}\right)\tilde{\Phi}\left(\mu_{n}+\frac{4}{\alpha_{n}};\mu_{n};\kappa_{n}\mu_{n}\right)}{\bar{\mathcal{A}}\Gamma\left(\frac{2}{\alpha_{n}}+\mu_{n}\right)^2\tilde{\Phi}\left(\mu_{n}+\frac{2}{\alpha_{n}};\mu_{n};\kappa_{n}\mu_{n}\right)^2}-1
\end{equation}\vspace{2mm}Likewise, for \textbf{Case II}, the second-order amount of fading can be derived using \eqref{n-th-moment-of-SNR-extreme} in \eqref{AF} resulting in\vspace{1.5mm}\begin{equation}\label{second-order-AF-Extreme}
  AF_{\gamma}^{e}=\dfrac{e^{2m_{n}}\Gamma\left(\frac{4}{\alpha_{n}}+1\right)\Phi\left(\frac{4}{\alpha_{n}}+1;2;2m_{n}\right)}{\bar{\mathcal{A}}^e2m_{n}\Gamma\left(\frac{2}{\alpha_{n}}+1\right)^2\Phi\left(\frac{2}{\alpha_{n}}+1;2;2m_{n}\right)^2}-1
\end{equation}
\subsection{Outage probability}
The OP is defined as the probability that the SNR per symbol falls below a certain threshold $\gamma_{th}$. By employing \eqref{CDF-of-system-SNR-f-final} and \eqref{CDF-of-system-SNR-f-final-extreme}, this probability can be written for \textbf{Case I} as\vspace{1.5mm}\begin{equation}\label{OutageProbability}
\begin{aligned}
  P_{out}&=P\left(\gamma\leq \gamma_{th} \right)\\
  &=1-\bar{\mathcal{A}}Q_{\mu_{n}}\left(\sqrt{2\kappa_{n}\mu_{n}},\sqrt{2\sigma_{n} \gamma_{th}^{\alpha_{n}/2}} \right).
  \end{aligned}
\end{equation}\vspace{1.5mm}Similarly, for \textbf{Case II}, we have that OP can be written as\vspace{1.5mm}\begin{equation}\label{OutageProbability-Extreme}
  P_{out}^{e}=1-\bar{\mathcal{A}}^eQ_{0}\left(2\sqrt{m_{n}},2\sqrt{m_{n}(\gamma_{th}/\bar{\gamma})^{\alpha_{n}/2} }\right).
\end{equation}

\subsection{Bit Error Rate}
\subsubsection{Coherent Modulation Schemes}
For a given SNR $\gamma$, the BER for a variety of coherent modulation schemes can be obtained by averaging the PDF $f_{\gamma}$ over the conditional additive white Gaussian noise (AWGN) BER given as {\cite[Eq. 17]{badarneh2016performance}}\vspace{1.5mm}\begin{equation}\label{BER-definition}
  P_{b}(e|\gamma)\triangleq(\phi/2)\mbox{erfc}\left(\rho\sqrt{\gamma/2}\right)
\end{equation}\vspace{1.5mm}where $\mbox{erfc}(\cdot)$ is the complementary error function {\cite[Eq. 8.250(4)]{gradshteyn2014table}}. The constants $\phi$ and $\rho$ vary depending on the type of modulation scheme. For BFSK, $\phi=1$ and $\rho=1.$ For BPSK, $\phi=1$ and $\rho=\sqrt{2}.$ For 4-QAM and QPSK, $\phi=2$ and $\rho=1.$ Finally, for M-PAM, $\phi=2(1-1/M)$ and $\rho=\sqrt{6/(M^2-1)}.$ Thus, for \textbf{Case I} we average \eqref{pdf-of-SNR-final} over \eqref{BER-definition} to obtain the integral\vspace{1.5mm}\begin{equation}\label{BER}
  P_{b}=\dfrac{\phi}{2}\int_{0}^{\infty}p_{\gamma}(\gamma)\mbox{erfc}\left(\rho\sqrt{\gamma/2}\right)d\gamma.
\end{equation} Substituting \eqref{pdf-of-SNR-final} into \eqref{BER}, one has that \begin{equation}\label{BER-main-int-non-co}
  P_{b}=\dfrac{\phi\mathcal{A}}{2}\int_{0}^{\infty}\delta(\gamma)\mbox{erfc}\left(\rho\sqrt{\gamma/2}\right)d\gamma+\dfrac{\phi\mathcal{K}_{n}\bar{\mathcal{A}}}{2}\mathcal{I}_{2}
\end{equation}\vspace{1.5mm}where\vspace{1.5mm}\begin{equation}
\mathcal{I}_{2}=\int_{0}^{\infty}\dfrac{\gamma^{(\alpha_{n}/2)\omega_{n}-1}}{e^{\sigma_{n}\gamma^{\alpha_{n}/2}}}I_{\mu_{n}-1}\left(\sqrt{\theta_{n}\gamma^{\frac{\alpha_{n}}{2}}}\right)\mbox{erfc}\left(\rho\sqrt{\frac{\gamma}{2}}\right)d\gamma.
\end{equation}\vspace{1.5mm}Following the same method as in Section \ref{sec:2}, using {\cite[Eq. 8.4.14(2)]{brychkov1986integrals}} to express the complementary error function in terms of its Mellin-Barnes representation, we obtain after some simple mathematical manipulations the BER under coherent modulation schemes for \textbf{Case I} and \textbf{Case II} as shown in \eqref{final-BER-Extreme} and \eqref{final-BER-Extreme} of Table \ref{table:2}, respectively. \begin{table}\fontsize{12pt}{12pt}
  \caption{BER UNDER COHERENT MODULATION}\label{table:2}
  \centering
  \begin{minipage}{1.00\columnwidth}
  \hrule
\hrule
\smallskip
\vspace{1.5mm}\begin{equation}\label{final-BER}
\begin{aligned}
  P_{b}=&\dfrac{\phi}{2}\mathcal{A}+\bar{\mathcal{A}}\mathcal{K}_{n}\dfrac{\phi\sqrt{\pi}}{2}\left(\dfrac{(\sqrt{2})^{\alpha_{n}}}{\rho^{\alpha_{n}}}\right)^{\omega_{n}}\\
       &\times H_{2,1:0,1;1,3}^{0,2:1,0;1,0}\left(\left.\begin{array}{c}
\sigma_{n}\dfrac{(\sqrt{2})^{\alpha_{n}}}{\rho^{\alpha_{n}}}
\\
\\ \theta_{n}\dfrac{(\sqrt{2})^{\alpha_{n}}}{4\rho^{\alpha_{n}}}
\end{array}\right|\begin{array}{c}
\boldsymbol{\Phi}(\alpha_{n},\mu_{n},\omega_{n})
\\
\\ \boldsymbol{\Psi}(\alpha_{n},\mu_{n},\omega_{n})
\end{array}\right)
\end{aligned}
\end{equation}\vspace{1.5mm}\begin{equation}\label{final-BER-Extreme}
\begin{aligned}
  P_{b}^{e}=&\dfrac{\phi}{2}\mathcal{A}^e+a_{n}\bar{\mathcal{A}}^e\dfrac{\phi \sqrt{\pi}}{2}\left(\dfrac{(\sqrt{2})^{\alpha_{n}}}{\rho^{\alpha_{n}}}\right)^{\frac{1}{2}}\\
  &\times H_{2,1:0,1;1,3}^{0,2:1,0;1,0}\left(\left.\begin{array}{c}
b_{n}\dfrac{(\sqrt{2})^{\alpha_{n}}}{\rho^{\alpha_{n}}}
\\
\\ c_{n}\dfrac{(\sqrt{2})^{\alpha_{n}}}{4\rho^{\alpha_{n}}}
\end{array}\right|\begin{array}{c}
\boldsymbol{\Phi}\left(\alpha_{n},2,\dfrac{1}{2}\right)
\\
\\ \boldsymbol{\Psi}\left(\alpha_{n},2,\dfrac{1}{2}\right)
\end{array}\right).
\end{aligned}
\end{equation}\vspace{1.5mm}\begin{equation*}
  \begin{aligned}
    \boldsymbol{\Phi}&\equiv\left(1-\frac{\alpha_{n}}{2}\omega_{n};\frac{\alpha_{n}}{2},\frac{\alpha_{n}}{2} \right),\left(\frac{1}{2}-\frac{\alpha_{n}}{2}\omega_{n};\frac{\alpha_{n}}{2},\frac{\alpha_{n}}{2} \right):\; - \; ;\left(\frac{\mu_{n}}{2},1\right)\\
    \boldsymbol{\Psi}&\equiv\left(-\frac{\alpha_{n}}{2}\omega_{n};\frac{\alpha_{n}}{2},\frac{\alpha_{n}}{2} \right)\; :\left(0,1\right) ; \; (\omega_{n}-1,1),(1-\omega_{n},1),\left(\frac{\mu_{n}}{2},1\right)
  \end{aligned}
  \end{equation*}\vspace{1.5mm}
  \hrule
\hrule
  \end{minipage}
\end{table}

\subsubsection{Non-Coherent Modulation Schemes}
Under a range of different non-coherent modulation schemes, the BER of the end-to-end SNR $\gamma$ can be written as {\cite[Eq. 27]{badarneh2016performance}}\vspace{1.5mm}\begin{equation}\label{BER-Non-oherent-definition}
  P_{bn}\triangleq\phi\int_{0}^{\infty}e^{-\rho\gamma}f_{\gamma}(\gamma)d\gamma.
\end{equation}\vspace{1.5mm}The values of $\phi$ and $\rho$ also vary depending on the modulation scheme. Specifically, for BFSK, $\phi=\frac{1}{2}$ and $\rho=\frac{1}{2}.$ For DBPSK, $\phi=\frac{1}{2}$ and $\rho=1$ and for $M$-FSK $\phi=(M-1)/2$ and $\rho=\frac{1}{2}.$ Upon substituting \eqref{pdf-of-SNR-final} into \eqref{BER-Non-oherent-definition}, it is clear that the resulting integral is of the form of the MGF integral given in \eqref{MGF-int}. It is therefore straightforward to show that\vspace{1.5mm}\begin{equation}\label{BER-non-coherent-final}
  P_{bn}=\phi\mathcal{M}_{\gamma}(\rho).
\end{equation}\vspace{1.5mm}Similarly, the BER for non-coherent modulation schemes under \textbf{Case II}, can be derived using \eqref{Final-MGF-Extreme} resulting in \vspace{1.5mm}\begin{equation}\label{BER-non-coherent-final-Extreme}
 P_{bn}^{e}=P_{bn}=\phi\mathcal{M}_{\gamma}^{e}(\rho).
\end{equation}

\subsection{Channel Capacity}
In this section, we investigate the capacity of $\alpha-\kappa-\mu$ and $\alpha-\kappa-\mu$-Extreme fading channels under various adaptive transmission schemes.

\subsubsection{Optimal Rate Adaptation}
 The channel capacity of the received ene-to-end SNR $\gamma$ with density function $f_{\gamma}(\gamma)$ under ORA (also known as ergodic capacity) is obtained by averaging the capacity of an additive AWGN $\boldsymbol{\mathcal{C}}_{O}=B\log_{2}(1+\gamma)$ over $f_{\gamma}(\gamma),$ where $B$ is the bandwidth of the channel. For a $n$-hop DF-based multi-hop system, we have\vspace{1.5mm}\begin{equation}\label{ErgodicDefn}
  \boldsymbol{\mathcal{C}}_{O}=\dfrac{B}{n\ln2}\int_{0}^{\infty}\ln(1+\gamma)f_{\gamma}(\gamma)d\gamma.
\end{equation}\vspace{1.5mm}For \textbf{Case I}, we substitute \eqref{pdf-of-SNR-final} into \eqref{ErgodicDefn} and set $y=\gamma^{\alpha_{n}/2}$ to obtain\vspace{1.5mm}\begin{equation}
 \dfrac{n\boldsymbol{\mathcal{C}}_{O}}{B}= \dfrac{\bar{\mathcal{A}}\alpha_{n}\mathcal{K}_{n}}{\ln4}\overset{\mathcal{I}_{3}}{\overbrace{\int_{0}^{\infty}\dfrac{\ln\left( 1+y^{\frac{2}{\alpha_{n}}}\right)}{y^{1-\omega_{n}}e^{\sigma_{n}y}}I_{\mu_{n}-1}\left(\sqrt{\theta_{n}y}\right)dy}}
\end{equation}\vspace{1.5mm}Proceeding as before, making routine use of the definition of the Fox H-function of a single variable from {\cite[Eq. 1.1.1]{mathai1978h}} yields the double Mellin-Barnes integral\vspace{1.5mm}\begin{equation}\label{ergodic-int-final}
\begin{aligned}
  \mathcal{I}_{3}=&\dfrac{1}{(2\pi i)^2}\int_{\mathcal{L}_{r}}\int_{\mathcal{L}_{t}}\dfrac{\Gamma\left[ \omega_{n}+\left(\frac{2}{\alpha_{n}}\right)r+t\right]\Gamma(\omega_{n}-1-t)}{\Gamma\left( 1-\frac{\mu_{n}}{2}+t\right)\Gamma\left(\frac{\mu_{n}}{2}-t\right)}\\
&\times\dfrac{\pi\Gamma(1-r)\Gamma(r)\Gamma(r)\left(\frac{1}{\sigma_{n}}\right)^{\frac{2r}{\alpha_{n}}}\left(\frac{\theta_{n}}{4\sigma_{n}}\right)^{t}drdt}{\sigma_{n}^{\omega_{n}}\Gamma(1+r)\Gamma(\omega_{n}+t)\Gamma\left(\frac{\mu_{n}}{2}-t\right)}
\end{aligned}
\end{equation}\vspace{1.5mm}where $Re\left(\omega_{n}+2r/\alpha_{n}+t\right)>0.$ Comparing \eqref{ergodic-int-final} with {\cite[Eq. 2.57]{mathai2009h}} results i the closed-form expression for $\boldsymbol{\mathcal{C}}_{O}$ under \textbf{Case I} as shown in \eqref{Ergodic-Capacity-Final} of Table \ref{table:3}. Using Remark \ref{remark:1}, we obtain by inspection from \eqref{Ergodic-Capacity-Final} the ORA capacity for \textbf{Case II} as in \eqref{Ergodic-Capacity-Extreme-Final} of Table \ref{table:3}.
\begin{table}\fontsize{12pt}{12pt}
\caption{ORA CHANNEL CAPACITY}\label{table:3}
  \centering
  \begin{minipage}{1.00\columnwidth}
  \hrule
\hrule
\vspace{1.5mm}\begin{equation}\label{Ergodic-Capacity-Final}
  \boldsymbol{\mathcal{C}}_{O}= \dfrac{B\bar{\mathcal{A}}\pi\alpha_{n}\mathcal{K}_{n}}{2n\sigma_{n}^{\omega_{n}}\ln2}H_{1,0:2,2;1,3}^{0,1:1,2;1,0}\left(\left.\begin{array}{c}
\left(\dfrac{1}{\sigma_{n}} \right)^{\frac{2}{\alpha_{n}}}
\\
\\ \dfrac{\theta_{n}}{4\sigma_{n}}
\end{array}\right|\begin{array}{c}
\boldsymbol{\Omega}(\alpha_{n},\mu_{n},\omega_{n})
\\
\\  \boldsymbol{\Pi}(\alpha_{n},\mu_{n},\omega_{n})
\end{array}\right)
\end{equation}\vspace{1.5mm}\begin{equation}\label{Ergodic-Capacity-Extreme-Final}
  \boldsymbol{\mathcal{C}}_{O}^e= \dfrac{B\bar{\mathcal{A}}^e\pi\alpha_{n}a_{n}}{2n\sqrt{b_{n}}\ln2}H_{1,0:2,2;1,3}^{0,1:1,2;1,0}\left(\left.\begin{array}{c}
\left(\dfrac{1}{b_{n}} \right)^{\frac{2}{\alpha_{n}}}
\\
\\ \dfrac{c_{n}}{4b_{n}}
\end{array}\right|\begin{array}{c}
\boldsymbol{\Omega}\left(\alpha_{n},2,\dfrac{1}{2}\right)
\\
\\ \boldsymbol{\Pi}\left(\alpha_{n},2,\dfrac{1}{2}\right)
\end{array}\right)
\end{equation}\vspace{1.5mm}\begin{equation*}
\begin{aligned}
  \boldsymbol{\Omega}&\equiv\left(1-\omega_{n};\dfrac{2}{\alpha_{n}},1\right):(1,1),(1,1);\left(\dfrac{\mu_{n}}{2},1\right)\\
  \boldsymbol{\Pi}&\equiv -:(1,1),(0,1);(\omega_{n}-1,1),(1-\omega_{n},1),\left(\dfrac{\mu_{n}}{2},1\right)
\end{aligned}
\end{equation*}\vspace{1.5mm}
\hrule
\hrule
\end{minipage}
\end{table} Using integration by parts, \cite{annamalai2010estimating} showed that \eqref{ErgodicDefn} has the alternative representation\vspace{1.5mm}\begin{equation}\label{ErgodicDefn-II}
  \boldsymbol{\mathcal{C}}_{O}=\dfrac{B}{n\ln2}\int_{0}^{\infty}\dfrac{\left[1-F_{\gamma}(\gamma)\right]}{1+\gamma}d\gamma.
\end{equation}\vspace{1.5mm}Therefore, substituting \eqref{CDF-of-system-SNR-f-final} and \eqref{CDF-of-system-SNR-f-final-extreme} into \eqref{ErgodicDefn-II} yields expressions for channel capacity with ORA under \textbf{Case I} and \textbf{Case II} respectively as \vspace{1.5mm}\begin{equation}\label{ErgodicI-AltI}
\boldsymbol{\mathcal{C}}_{O}=\dfrac{B\bar{\mathcal{A}}}{n\ln2}\int_{0}^{\infty}Q_{\mu_{n}}\left(\sqrt{2\kappa_{n}\mu_{n}},\sqrt{2\sigma_{n} \gamma^{\frac{\alpha_{n}}{2}}} \right)/(1+\gamma)d\gamma,
\end{equation}\vspace{1.5mm}\begin{equation}\label{ErgodicI-AltII}
\boldsymbol{\mathcal{C}}_{O}^{e}=\dfrac{B\bar{\mathcal{A}}^e}{n\ln2}\int_{0}^{\infty}Q_{0}\left(2\sqrt{m_{n}},2\sqrt{m_{n}(\gamma/\bar{\gamma})^{\frac{\alpha_{n}}{2}}}\right)/(1+\gamma)d\gamma.
\end{equation}

\subsubsection{Optimal Power and Rate Adaptation}

Under OPRA, the channel capacity of the end-to-end SNR $\gamma$ of a $n$-link DF-based multi-hop system with density $f_{\gamma}$ is defined as {\cite[Eq. 10]{annamalai2010estimating}}\vspace{1.5mm}\begin{equation}\label{OPRA-defn-I}
\begin{aligned}
  \boldsymbol{\mathcal{C}}_{P}\triangleq&\dfrac{B}{n}\int_{\gamma_{o}}^{\infty}\log_{2}\left(\dfrac{\gamma}{\gamma_{o}}\right)f_{\gamma}(\gamma)d\gamma\\
  =&\dfrac{\gamma_{o}B}{n\ln2}\int_{1}^{\infty}\ln(y)f_{\gamma}(\gamma_{o}y)dy, \mbox{upon setting }\gamma=\gamma_{o}y.
\end{aligned}
\end{equation}\vspace{1.5mm}The optimum cutoff $\gamma_{o}$ must satisfy the following relation \vspace{1.5mm}\begin{equation}\label{OPRA-cutoff-SNR-relation-I}
\int_{\gamma_{o}}^{\infty}\left(1/\gamma_{o}-1/\gamma\right)f_{\gamma}(\gamma)d\gamma=1.
\end{equation}\vspace{1.5mm}In order to evaluate \eqref{OPRA-defn-I} for \textbf{Case I}, we express the PDF \eqref{pdf-of-SNR-final} in terms of the Fox H-function as was done in Section \ref{sec:2} to obtain after some simple manipulations the integral\vspace{1.5mm}\begin{equation}
\begin{aligned}
  \boldsymbol{\mathcal{C}}_{P}=&\mathcal{K}^*\dfrac{1}{(2\pi i)^2}\int_{L_{r}}\int_{L_{t}}\overset{\mathcal{H}}{\overbrace{\int_{1}^{\infty}\dfrac{\ln\left(y^{\frac{2}{\alpha_{n}}}\right)}{y^{1-\omega_{n}-r-t}}dy}}\\
  &\times\dfrac{\Gamma(-r)\Gamma(\omega_{n}-1-t)\left(\sigma_{n}\gamma_{o}^{\frac{\alpha_{n}}{2}}\right)^{r}\left( \frac{\theta_{n}}{4}\gamma_{o}^{\frac{\alpha_{n}}{2}}\right)^t drdt}{\Gamma(\omega_{n}+t)\Gamma\left(1-\frac{\mu_{n}}{2}+t\right)\Gamma\left(\frac{\mu_{n}}{2}-t\right)}
\end{aligned}
\end{equation}\vspace{1.5mm}where $\mathcal{K}^*=\left(2\gamma_{o}^{\frac{\alpha_{n}}{2}\omega_{n}}\pi B\mathcal{K}_{n}\right)/(\alpha_{n}n\ln2\bar{\mathcal{A}}^{-1}).$  The integral $\mathcal{H}$ can be evaluated using integration by parts as $\mathcal{H}=2/\left[\alpha_{n}(\omega_{n}+r+t)^2\right]$ resulting in the double Mellin-Barnes contour integral\vspace{1.5mm}\begin{equation}\label{OPRA-Contour}
\begin{aligned}
  \boldsymbol{\mathcal{C}}_{P}=&\dfrac{2\mathcal{K}^*}{\alpha_{n}}\dfrac{1}{(2\pi i)^2}\int_{L_{r}}\int_{L_{t}}\dfrac{\Gamma(-r)\Gamma(\omega_{n}+r+t)^2 }{\Gamma(1+\omega_{n}+r+t)^2}\\
  &\times \dfrac{\Gamma(\omega_{n}-1-t)\left(\sigma_{n}\gamma_{o}^{\frac{\alpha_{n}}{2}}\right)^{r}\left( \frac{\theta_{n}}{4}\gamma_{o}^{\frac{\alpha_{n}}{2}}\right)^t drdt}{\Gamma(\omega_{n}+t)\Gamma\left(1-\frac{\mu_{n}}{2}+t\right)\Gamma\left(\frac{\mu_{n}}{2}-t\right)}
\end{aligned}
\end{equation}\vspace{1.5mm}where $\Re(\omega_{n}+r+t)<0.$ Finally, comparing \eqref{OPRA-Contour} with {\cite[Eq. 2.57]{mathai2009h}} gives a closed-form expression for the OPRA channel capacity  under \textbf{Case I} and \textbf{Case II} as shown in \eqref{Final-OPRA-capacity} and \eqref{Final-OPRA-capacity-II} of of Table \ref{table:4}, respectively.
\begin{table}\fontsize{12pt}{12pt}
\caption{OPRA CHANNEL CAPACITY}\label{table:4}
 \centering
 \begin{minipage}{1.00\columnwidth}
 \hrule
\hrule
\vspace{1.5mm}\begin{equation}\label{Final-OPRA-capacity}
\boldsymbol{\mathcal{C}}_{P}=\dfrac{4\gamma_{o}^{\frac{\alpha_{n}}{2}\omega_{n}}\pi B\mathcal{K}_{n}}{\alpha_{n}^{2}n\ln2\bar{\mathcal{A}}^{-1}}G_{2,2:0,1;1,3}^{0,2:1,0;1,0}\left(\left.\begin{array}{c}
\sigma_{n}\gamma_{o}^{\frac{\alpha_{n}}{2}}
\\
\\ \dfrac{\theta_{n}}{4}\gamma_{o}^{\frac{\alpha_{n}}{2}}
\end{array}\right|\begin{array}{c}
\boldsymbol{\spadesuit}(\mu_{n},\omega_{n})
\\
\\ \boldsymbol{\clubsuit}(\mu_{n},\omega_{n})
\end{array}\right)
\end{equation}\vspace{1.5mm}\begin{equation}\label{Final-OPRA-capacity-II}
  \boldsymbol{\mathcal{C}}_{P}^{e}=\dfrac{4\pi\bar{\mathcal{A}}^e\gamma_{o}^{\frac{\alpha_{n}}{4}} Ba_{n}}{\alpha_{n}^{2}n\ln2}G_{2,2:0,1;1,3}^{0,2:1,0;1,0}\left(\left.\begin{array}{c}
b_{n}\gamma_{o}^{\frac{\alpha_{n}}{2}}
\\
\\ \dfrac{c_{n}}{4}\gamma_{o}^{\frac{\alpha_{n}}{2}}
\end{array}\right|\begin{array}{c}
\boldsymbol{\spadesuit}\left(2,\dfrac{1}{2}\right)
\\
\\ \boldsymbol{\clubsuit}\left(2,\dfrac{1}{2}\right)
\end{array}\right)
\end{equation}\vspace{1.5mm}\begin{equation*}
\begin{aligned}
  \boldsymbol{\spadesuit}&\equiv(1-\omega_{n};1,1),(1-\omega_{n};1,1):-;\left(\dfrac{\mu_{n}}{2},1\right)\\
  \boldsymbol{\clubsuit}&\equiv(-\omega_{n};1,1);(-\omega_{n};1,1):(0,1);(\omega_{n}-1,1),(1-\omega_{n},1),\left(\dfrac{\mu_{n}}{2},1\right)
\end{aligned}
\end{equation*}\vspace{1.5mm}
\hrule
\hrule
\end{minipage}
\end{table} Applying integration by parts in \eqref{OPRA-cutoff-SNR-relation-I} results in the alternative representation for the cutoff SNR relation as {\cite[Eq. 13]{annamalai2010estimating}} \vspace{1.5mm}\begin{equation}\label{OPRA-cutoff-SNR-relation-II}
\begin{aligned}
  &\left[1-F_{\gamma}(\gamma_{o})\right]/\gamma_{o}-\int_{\gamma_{o}}^{\infty}f_{\gamma}(\gamma)/\gamma d\gamma=1\\
  \Longrightarrow&\;\gamma_{o}=\dfrac{1-F_{\gamma}(\gamma_{o})}{1+\int_{\gamma_{o}}^{\infty}f_{\gamma}(\gamma)/\gamma d\gamma}=\xi(\gamma_{o}).
\end{aligned}
\end{equation}\vspace{1.5mm}It is clear that the solution to \eqref{OPRA-cutoff-SNR-relation-II} is a fixed point of the non-linear equation $\gamma_{o}=\xi(\gamma_{o}).$ According to \cite{annamalai2010estimating}, $\gamma_{o}$ will always lie in the interval $[0,1]$ regardless of the fading model and number of relays used for the wireless system. The explanation for the latter relies on the properties of the CDF of $\gamma$ and is fully documented in Section 3 of \cite{annamalai2010estimating}. Thus, choosing a starting point in $[0,1],$ $\gamma_{o}$ can be computed via iterative schemes such as the Newton-Raphson method as\vspace{1.5mm}\begin{equation*}
  \gamma_{o}^{i}=\xi(\gamma_{o}^{i})
\end{equation*}\vspace{1.5mm}where $\gamma_{o}^{i}$ is the $i$-th iterate of $\gamma_{o}.$ The iterative procedure will be stopped when $|\gamma_{o}^{i+1}-\gamma_{o}^{i}|$ is sufficiently small. We also note again that using integration by parts on \eqref{OPRA-defn-I} yields {\cite[Eq. 11]{annamalai2010estimating}}
\vspace{1.5mm}\begin{equation}\label{OPRA-defn-II}
  \boldsymbol{\mathcal{C}}_{P}=\dfrac{B}{n\ln2}\int_{\gamma_{o}}^{\infty}\left[1-F_{\gamma}(\gamma)\right]/\gamma d\gamma.
\end{equation}\vspace{1.5mm}Therefore, we can use \eqref{CDF-of-system-SNR-f-final} and \eqref{CDF-of-system-SNR-f-final-extreme} in \eqref{OPRA-defn-II} to obtain the OPRA channel capacities for \textbf{Case I} and \textbf{Case II} respectively as\vspace{1.5mm}\begin{equation}\label{Final-OPRA-Case-I-UsingCDF}
  \boldsymbol{\mathcal{C}}_{P}=\dfrac{B\bar{\mathcal{A}}}{n\ln2}\int_{\gamma_{o}}^{\infty}Q_{\mu_{n}}\left(\sqrt{2\kappa_{n}\mu_{n}},\sqrt{2\sigma_{n} \gamma^{\frac{\alpha_{n}}{2}}} \right))/\gamma d\gamma,
\end{equation}\vspace{1.5mm}\begin{equation}\label{Final-OPRA-Case-II-UsingCDF}
  \boldsymbol{\mathcal{C}}_{P}^e=\dfrac{B\bar{\mathcal{A}}^e}{n\ln2}\int_{\gamma_{o}}^{\infty}Q_{0}\left(2\sqrt{m_{n}},2\sqrt{m_{n}(\gamma/\bar{\gamma})^{\frac{\alpha_{n}}{2}} }\right)/\gamma d\gamma
\end{equation}\vspace{1.5mm}Numerical integration routines from standard packages can be employed to evaluate \eqref{Final-OPRA-Case-I-UsingCDF} and \eqref{Final-OPRA-Case-II-UsingCDF}.

\subsubsection{Channel Inversion With Fixed Rate}
 Under this technique, the channel capacity of the end-to-end SNR $\gamma$ of a $n$-link DF-based multi-hop system with density $f_{\gamma}$ is given by\vspace{1.5mm}\begin{equation}\label{CIFR}
  \boldsymbol{\mathcal{C}}_{c}\triangleq\dfrac{B}{n}\log_{2}\left(1+\dfrac{1}{\int_{0}^{\infty}\left(f_{\gamma}(\gamma)/\gamma\right)d\gamma} \right)
\end{equation}\vspace{1.5mm}Since the expression obtained using \eqref{n-th-moment-of-SNR} and \eqref{n-th-moment-of-SNR-extreme} are highly restrictive, we can obtain more general ones by employing a similar procedure to the one used in deriving the MGF function of Section \ref{sec:2}. We only express the bessel function in each SNR PDF in terms of the Meijer G-function and then reverse the order of integration to obtain after some simplifications the single Mellin-Barnes integral\vspace{1.5mm}\begin{equation}\label{CIFR-Contour-I}
\begin{aligned}
  \int_{0}^{\infty}\dfrac{p_{\gamma}(\gamma)}{\gamma} d\gamma=&\dfrac{2\pi\mathcal{K}_{n}\bar{\mathcal{A}}}{\alpha_{n}\sigma_{n}^{\omega_{n}-\frac{2}{\alpha_{n}}}}\dfrac{1}{2\pi i}\oint_{L_{r}}\dfrac{\Gamma\left(\omega_{n}-\frac{2}{\alpha_{n}}+r\right)}{\Gamma(\omega_{n}+r)}\\
  &\times\dfrac{\Gamma(\omega_{n}-1-r)\left(\frac{\theta_{n}}{4\sigma_{n}}\right)^rdr}{\Gamma\left( 1-\frac{\mu_{n}}{2}+r\right)\Gamma\left(\frac{\mu_{n}}{2}-r\right)}
\end{aligned}
\end{equation}\vspace{1.5mm}where $Re\left(\omega_{n}-\frac{2}{\alpha_{n}}+r\right)>0.$ Comparing the contour integral in \eqref{CIFR-Contour-I} with {\cite[Eq. 1.1]{mathai2009h}} yields the CIFR channel capacity for \textbf{Case I} as\vspace{1.5mm}\begin{equation}\label{CIFR-Final-Case-I}
\dfrac{n\boldsymbol{\mathcal{C}_{c}}}{B}=\log_{2}\left\{1+\left[\dfrac{2\pi\mathcal{K}_{n}\bar{\mathcal{A}}}{\alpha_{n}\sigma_{n}^{\omega_{n}-\frac{2}{\alpha_{n}}}}G_{2,3}^{1,1}\left(\frac{\theta_{n}}{4\sigma_{n}}\left|\begin{array}{c}
\boldsymbol{\blacktriangle}\\
\boldsymbol{\blacktriangledown}
\end{array}\right.\right)\right]^{-1}\right\},
\end{equation}\vspace{1.5mm}where $\boldsymbol{\blacktriangle}\equiv1+\frac{2}{\alpha_{n}}-\omega_{n},\frac{\mu_{n}}{2}$ and $\boldsymbol{\blacktriangledown}\equiv\omega_{n}-1,1-\omega_{n},\frac{\mu_{n}}{2}.$
In an analogous fashion, the CIFR channel capacity for \textbf{Case II} can be derived with the contour condition $Re\left(1/2-\frac{2}{\alpha_{n}}+r\right)>0$ as\vspace{1.5mm}\begin{equation}\label{CIFR-Final-Case-II}
\dfrac{n\boldsymbol{\mathcal{C}_{c}^{e}}}{B}=\log_{2}\left\{1+\left[\dfrac{2\pi a_{n}\bar{\mathcal{A}}^e}{\alpha_{n}b_{n}^{\frac{1}{2}-\frac{2}{\alpha_{n}}}}G_{2,3}^{1,1}\left(\frac{c_{n}}{4b_{n}}\left|\begin{array}{c}
\boldsymbol{\blacktriangle}^*\\
\boldsymbol{\blacktriangledown}^*
\end{array}\right.\right)\right]^{-1}\right\}
\end{equation} \vspace{1.5mm}where $\boldsymbol{\blacktriangle}^*\equiv\frac{1}{2}+\frac{2}{\alpha_{n}},1$ and $\boldsymbol{\blacktriangledown}^*\equiv-\frac{1}{2},\frac{1}{2},1$ The Meijer G-function has been implemented in MATHEMATICA$^\circledR$ and MATLAB$^\circledR.$

\subsubsection{Truncated Channel Inversion With Fixed Rate}\label{TruncInvSubSec}
The channel inversion technique with fixed rate may suffer large capacity penalties as compared to other techniques. To circumvent this shortcoming, a modified inversion which inverts the channel fading above a predetermined truncated fade $\gamma_{o}$ is often used and is defined for $n$-link DF-based multi-hop systems as\vspace{1.5mm}\begin{equation}\label{TIFR}
  \boldsymbol{\mathcal{C}}_{T}\triangleq\dfrac{B}{n}\log_{2}\left(1+\dfrac{1}{\int_{\gamma_{o}}^{\infty}f_{\gamma}(\gamma)/\gamma d\gamma} \right)Pr\left(\gamma>\gamma_{o}\right).
\end{equation}\vspace{1.5mm}Note that $\boldsymbol{\mathcal{C}}_{T}\rightarrow\boldsymbol{\mathcal{C}}_{c}$ from above as $\gamma_{o}\rightarrow 0.$ That is, the gap between $\boldsymbol{\mathcal{C}}_{T}$ and $\boldsymbol{\mathcal{C}}_{c}$ depends on $Pr\left(\gamma>\gamma_{o}\right)$ in \eqref{TIFR}. Making the substitution $y=\gamma/\gamma_{o},$ the integral in \eqref{TIFR} can be re-written as\vspace{1.5mm}\begin{equation}\label{TIFR-integral}
  \mathcal{J}=\int_{\gamma_{o}}^{\infty}f_{\gamma}(\gamma)/\gamma d\gamma=\int_{1}^{\infty}\dfrac{1}{y}f_{y}(\gamma_{o}y)dy
\end{equation}\vspace{1.5mm}Substituting \eqref{pdf-of-SNR-final} into \eqref{TIFR-integral} and following the simple procedure undertaken in deriving OPRA channel capacity yields \begin{table}\fontsize{12pt}{12pt}
\caption{TIFR CHANNEL CAPACITY INTEGRAL $\mathcal{J}$}\label{table:5}
  \centering
  \begin{minipage}{1.00\columnwidth}
  \hrule
\hrule
\vspace{1.5mm}\begin{equation}\label{Final-TIFR-integral_Case-I}
\mathcal{J}=\dfrac{-2\pi\bar{\mathcal{A}}\left(\gamma_{o}^{\frac{\alpha_{n}}{2}\omega_{n}} \mathcal{K}_{n}\right)}{\alpha_{n}\gamma_{o}}G_{1,2:0,1;1,3}^{0,1:1,0;1,0}\left(\left.\begin{array}{c}
\sigma_{n}\gamma_{o}^{\frac{\alpha_{n}}{2}}
\\
\\ \dfrac{\theta_{n}}{4}\gamma_{o}^{\frac{\alpha_{n}}{2}}
\end{array}\right|\begin{array}{c}
\boldsymbol{\spadesuit}^*(\mu_{n},\omega_{n})
\\
\\ \boldsymbol{\clubsuit}^*(\mu_{n},\omega_{n})
\end{array}\right)
\end{equation}\vspace{1.5mm}\begin{equation}\label{Final-TIFR-integral_Case-II}
  \mathcal{J}=\dfrac{-2\pi\bar{\mathcal{A}}^e\left(\gamma_{o}^{\frac{\alpha_{n}}{4}} a_{n}\right)}{\alpha_{n}}G_{1,1:0,1;1,3}^{0,1:1,0;1,0}\left(\left.\begin{array}{c}
b_{n}\gamma_{o}^{\frac{\alpha_{n}}{2}}
\\
\\ \dfrac{c_{n}}{4}\gamma_{o}^{\frac{\alpha_{n}}{2}}
\end{array}\right|\begin{array}{c}
\boldsymbol{\spadesuit}^*\left(2,\dfrac{1}{2}\right)
\\
\\ \boldsymbol{\clubsuit}^*\left(2,\dfrac{1}{2}\right)
\end{array}\right)
\end{equation}\vspace{1.5mm}\begin{equation*}
  \boldsymbol{\spadesuit}^*\equiv\left(1+\frac{2}{\alpha_{n}}-\omega_{n};1,1\right):-;\left(\dfrac{\mu_{n}}{2},1\right)\\
\end{equation*}\vspace{1.5mm}\begin{equation*}
\begin{aligned}
\boldsymbol{\clubsuit}^*\equiv&\left(\frac{2}{\alpha_{n}}-\omega_{n};1,1\right),\left(\frac{2}{\alpha_{n}}-\omega_{n};1,1\right):(0,1);(\omega_{n}-1,1),\\
&(1-\omega_{n},1),\left(\dfrac{\mu_{n}}{2},1\right)
\end{aligned}
\end{equation*}\vspace{1.5mm}
\hrule
\hrule
\end{minipage}
\end{table} The double Mellin-Barnes contour integral must satisfy $Re\left(\omega_{n}-\frac{2}{\alpha_{n}}+r+t\right)<0$. Therefore, putting \eqref{Final-TIFR-integral_Case-I} and \eqref{Final-TIFR-integral_Case-II} into \eqref{TIFR} gives the TIRF channel capacity for \textbf{Case I} and \textbf{Case II}, respectively. Alternatively, for \textbf{Case I}, another closed-form expression for the integral in \eqref{TIFR} can be obtained by making use of the consecutive substitutions $x=\gamma^{\alpha_{n}/2}$ followed by $\sqrt{x}=u/\sqrt{2\sigma_{n}}$ which yield\vspace{1.5mm}\begin{equation}\label{TIFR-CaseI-V2}
  \mathcal{J}=\dfrac{\bar{\mathcal{A}}(2\sigma_{n})^{\frac{2}{\alpha_{n}}}}{2(2\mu_{n}\kappa_{n})^{\omega_{n}-1}}\mathcal{Q}_{\mu^*,\mu_{n}-1}\left( \sqrt{2\mu_{n}\kappa_{n}},\sqrt{2\sigma_{n}\gamma_{o}^{\frac{\alpha_{n}}{2}}}\right)
\end{equation}\vspace{1.5mm}where $\mu^*=\mu_{n}-\frac{4}{\alpha_{n}},$ $\alpha_{n}\mu_{n}>4$ and $\mathcal{Q}_{M,N}(\cdot,\cdot)$ is the Nuttall Q-function defined as {\cite[Eq. 2]{sun2010monotonicity}}
 with $b,M,N\geq0$ and $a>0.$ A similar procedure can be followed for \textbf{Case II}. However, the resulting integral does not satisfy the $M\geq0$ condition and hence fails to attain a Nuttall Q-function representation. The Nuttall Q-function is a generalization of the Marcum Q-function. Specifically, when $M=N+1,$ $\mathcal{Q}_{N+1,N}(a,b)=a^NQ_{N+1}(a,b).$ The latter is clearly evident from \eqref{TIFR-CaseI-V2} in the case $\alpha=2.$

\section{BER and Capacity Approximations for Low SNR Regimes}\label{sec:4}
In this section, we explore the asymptotic behaviour of the BER by employing asymptotic Poincare series expansions for both the $\alpha-\kappa-\mu$ and $\alpha-\kappa-\mu$-Extreme fading models in \eqref{pdf-of-SNR-final} and \eqref{pdf-of-SNR-final-xtreme}. Following the idea from \cite{annamalai2014asymptotic} of using {\cite[Eq. 8.445]{gradshteyn2014table}} and {\cite[Eq. 0.316]{gradshteyn2014table}}, we have that the PDFs of the SNRs for the $n$-th hop of \textbf{Case I} and \textbf{Case II} fading channels have Poincare asymptotic series expansion near the origin given respectively by\vspace{1.5mm}\begin{equation}\label{Poincare-Series-Case-I}
  p_{\gamma}(\gamma)\;\dot{\thicksim}\;\mathcal{A}\delta(\gamma)+\bar{\mathcal{A}}\mathcal{K}_{n}\left(\dfrac{\theta_{n}}{4} \right)^{\omega_{n}-1}\smashoperator[r]{\sum_{k=0}^{N-1}}d_{k}\gamma^{t_{k}}+\mathcal{O}\left( \gamma^{t+N-1}\right)
\end{equation}\vspace{1.5mm}where $\mathcal{O}$ denotes the order of a term in the asymptotic series and\vspace{1.5mm}\begin{equation}\label{asymptotic-series-symbols-I}
  d_{k}=\smashoperator[r]{\sum_{j=0}^{k}}\dfrac{(-1)^j\sigma_{n}^{j}\left( \frac{\theta_{n}}{4}\right)^{k-j}}{j!(k-j)!\Gamma(k+\mu_{n}-j)},\;t_{k}=\alpha_{n}(k+\mu_{n})/2-1.
\end{equation}\vspace{1.5mm}\begin{equation}\label{Poincare-Series-Case-II}
  p_{\gamma}^{e}(\gamma)\;\dot{\thicksim}\;\mathcal{A}^e\delta(\gamma)+\bar{\mathcal{A}}^e\dfrac{a_{n}\sqrt{c_{n}}}{2}\smashoperator[r]{\sum_{k=0}^{N-1}}d_{k}^{e}\gamma^{t_{k}^{e}}+\mathcal{O}\left( \gamma^{t+N-1}\right)
\end{equation}\vspace{1.5mm}where\vspace{1.5mm}\begin{equation}\label{asymptotic-series-symbols-II}
  d_{k}^{e}=\smashoperator[r]{\sum_{j=0}^{k}}\dfrac{(-1)^jb_{n}^j\left( \frac{c_{n}}{4}\right)^{k-j}}{j!(k-j)!\Gamma(k+2-j)},\;t_{k}^{e}=\alpha_{n}(k+1)/2-1.
\end{equation}

\subsection{Asymptotic BER at low SNR levels}
We will now attempt to derive asymptotic BER expressions based on the series expansions in \eqref{Poincare-Series-Case-I} and \eqref{Poincare-Series-Case-II}.
\subsubsection{Coherent Modulation Schemes}
Substituting \eqref{Poincare-Series-Case-I} and \eqref{Poincare-Series-Case-II} into \eqref{BER-definition} and employing {\cite[Eq. 8.4.15(2)]{brychkov1986integrals}}, we have that the asymptotic BER under coherent modulation for \textbf{Case I} and \textbf{Case II} are given respectively by\vspace{1.5mm}\begin{equation}\label{BER-Case-I-Asymp}
  \mathcal{P}_{b}\;\dot{\thicksim}\;\dfrac{\phi}{2}\mathcal{A}+\bar{\mathcal{A}}\mathcal{K}_{n}\left(\dfrac{\theta_{n}}{4} \right)^{\omega_{n}-1}\dfrac{\phi}{\sqrt{\pi}}\smashoperator[r]{\sum_{k=0}^{N-1}}d_{k}\dfrac{2^{t_{k}}\Gamma\left(\frac{3}{2}+t_{k}\right)}{(t_{k}+1)\rho^{2(t_{k}+1)}},
\end{equation}\vspace{1.5mm}\begin{equation}\label{BER-Case-II-Asymp}
  \mathcal{P}_{b}^{e}\;\dot{\thicksim}\;\dfrac{\phi}{2}\mathcal{A}^e+\phi\bar{\mathcal{A}}^e\dfrac{a_{n}\sqrt{c_{n}}}{2\sqrt{\pi}}\smashoperator[r]{\sum_{k=0}^{N-1}}d_{k}^{e}\dfrac{2^{t_{k}^{e}}\Gamma\left(\frac{3}{2}+t_{k}^{e}\right)}{(t_{k}^{e}+1)\rho^{2(t_{k}^{e}+1)}}.
\end{equation}
\subsubsection{Non-coherent Modulation Schemes}
Similarly, by substituting \eqref{Poincare-Series-Case-I} and \eqref{Poincare-Series-Case-II} into \eqref{BER-Non-oherent-definition} and employing {\cite[Eq. 8.4.3(1)]{brychkov1986integrals}}, we have that the asymptotic BER under non-coherent modulation for \textbf{Case I} and \textbf{Case II} are given respectively by\vspace{1.5mm}\begin{equation}\label{BER-Case-I-Asymp-NonCo}
  \mathcal{P}_{bn}\;\dot{\thicksim}\;\phi\mathcal{A}+\phi\bar{\mathcal{A}}\mathcal{K}_{n}\left(\dfrac{\theta_{n}}{4} \right)^{\omega_{n}-1}\smashoperator[r]{\sum_{k=0}^{N-1}}d_{k}\dfrac{\Gamma(t_{k}+1)}{\rho^{t_{k}+1}}
  \end{equation}\vspace{1.5mm}\begin{equation}\label{BER-Case-II-Asymp-NonCo}
  \mathcal{P}_{bn}^{e}\;\dot{\thicksim}\;\phi\mathcal{A}^e+\phi\bar{\mathcal{A}}^e\dfrac{a_{n}\sqrt{c_{n}}}{2}\smashoperator[r]{\sum_{k=0}^{N-1}}d_{k}^{e}\dfrac{\Gamma(t_{k}^{e}+1)}{\rho^{t_{k}^{e}+1}}.
\end{equation}

\section{Numerical Experiments and Simulations}\label{sec:5}
 In this section we have presented the graphical results of the performance metrics derived in the preceding sections. All computations are carried out in MATLAB$^\circledR$(version 2014a).Nonetheless, based on linear optimization, \cite{chergui2018rician} recently proposed versatile GPU-enabled MATLAB$^\circledR$ and $C/MEX$ codes with automated contour computation for the general multivariate Fox H-function of the type in \cite{mathai2009h} and we use a modified version of it in the present work.  
Figure \ref{fig:1} and Figure \ref{fig:2} present the AF vs average SNR under severe and moderate fading conditions for both \textbf{Case I} and \textbf{Case II}, respectively. It is evident that AF decreases with increments in values for any of the fading parameters of both models.  This makes sense since the parameter $m$ is inversely proportional to the fading severity. We also note that both models perform almost identically for moderate fading with the extreme distribution showing a slightly lower AF. Furthermore, AF under \textbf{Case II} stabilizes to a constant as SNR decreases indefinitely. While increasing the number of hops appears to lower the AF for moderate fading conditions, for severe fading, there is a steep increase in AF at low SNR with increases in signal power, the number of multi-clusters $\mu$ or $m$ having little to no positive effect. \begin{figure}[!htb]
\centering\includegraphics[clip, trim=0.25 6cm 0.25cm 6cm,width=21pc]{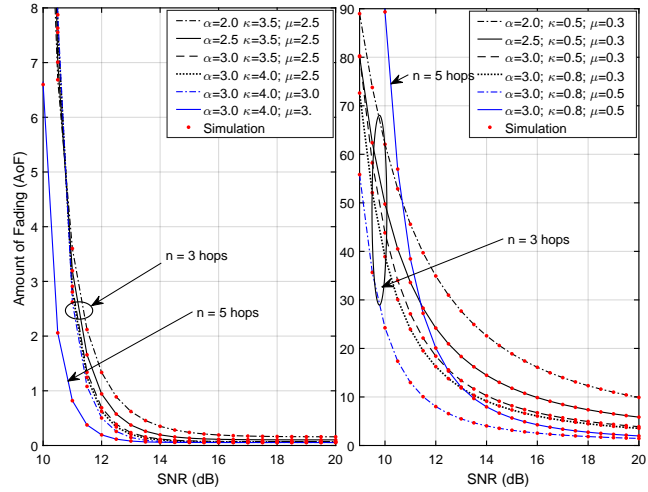}
  \caption{AF curves for the $\alpha-\kappa-\mu$ case showing the effects of variation in model parameter values and hop numbers under moderate and severe fading conditions.}\label{fig:1}
\end{figure} \begin{figure}[!htb]
\centering\includegraphics[clip, trim=0.25 6cm 0.25cm 6cm,width=21pc]{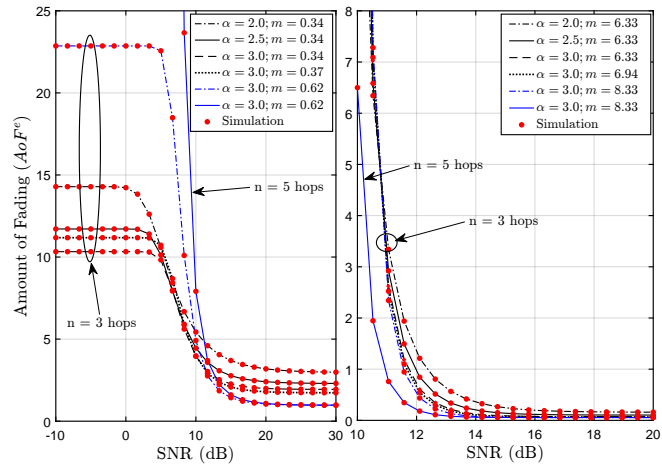}
  \caption{AF curves for the $\alpha-\kappa-\mu$-Extreme case showing the effects of variation in model parameter values and hop numbers under moderate and severe fading conditions similar to \textbf{Case I}.}\label{fig:2}
\end{figure} { Figure \ref{fig:3} illustrates the OP vs average SNR curves for \textbf{Case I} under moderate fading conditions. From the figure, we can deduce that increasing any of the fading parameters values results in notable system performance enhancement with steeper increases noted for increments in $\mu,$ the multi-cluster parameter. Increasing the number of hops also improves system performance, especially at lower SNR levels which is a particularly desirable trait. The increase in performance is albeit of a diminishing type. This implies that there is an optimal number/range of hops that maximizes system performance. That is, relaying for the sole purpose of improving outage is only fruitful up to a certain number of relays upon which any additional relaying is futile. \begin{figure}[!htb]
\centering\includegraphics[clip, trim=0.2 5.5cm 0.2cm 5.5cm,width=21pc]{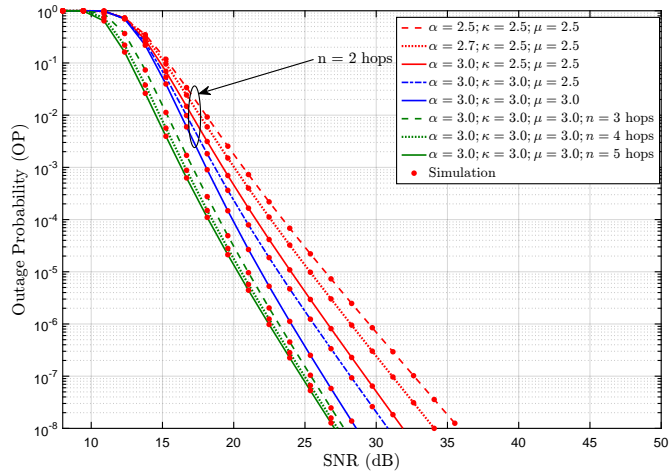}
  \caption{OP curves for the $\alpha-\kappa-\mu$ case under moderate fading showing the effects of increments in hop number and fading model parameter values.}\label{fig:3}
\end{figure}
Figure \ref{fig:6} depicts the behaviour of BER under non-coherent DBPSK modulation scheme for \textbf{Case I} in moderate fading conditions. Consistent with observations from AF and OP analyses, there is notable system improvement with increments in fading model parameter values and a diminishing system performance improvement with hop number increase at low SNR levels. An additional system behavior is observed for high SNR level which shows that there is a non-null BER even as signal power increases indefinitely. This characteristic was observed by Rabelo et al. \cite{rabelo2009mu} for the linear $\kappa-\mu$-Extreme fading model where they showed that as SNR increases, BER under non-coherent DPSK modulation approaches the constant $(1/2)\exp(-2m).$  \begin{figure}[!htb]
  \centering\includegraphics[clip, trim=0.25 6cm 0.25cm 6cm,width=21pc]{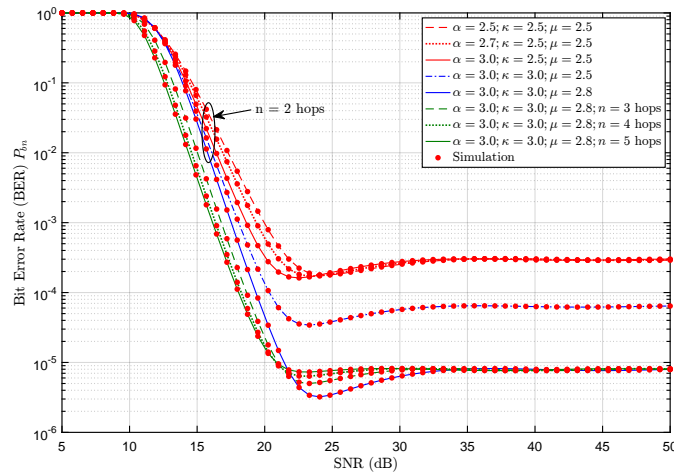}
  \caption{BER curves for the $\alpha-\kappa-\mu$ case under non-coherent DBPSK modulation showing the effects of increasing model parameters and the number of hops and $N$ in Poincare approximations for moderate fading conditions}\label{fig:6}
\end{figure} \begin{figure}[htbp]
  \centering\includegraphics[clip, trim=0.2 5.5cm 0.2cm 5.5cm,width=21pc]{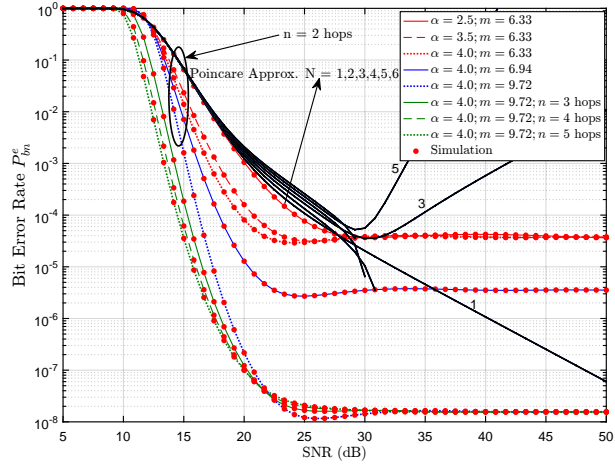}
  \caption{BER curves for the $\alpha-\kappa-\mu$-Extreme case under non-coherent DBPSK modulation showing the effects of increasing model parameters, the number of hops and $N$ in Poincare approximations for moderate fading conditions}\label{fig:7}
\end{figure}

 Under similar fading conditions of \textbf{Case I}, Figure \ref{fig:7} presents BER curves under non-coherent DBPSK modulation for \textbf{Case II}. We once more observe a consistent behaviour in terms of parameter and hop number variation. Furthermore, BER levels are considerably lower as compared to \textbf{Case I} even under moderate fading severity.   Figure \ref{fig:7} also shows that Poincare series-based asymptotic BER approximations perform quite well at low SNR regimes for the choice in parameter values, but that an increase in $N$ does not seem to improve convergence or guarantee improvement in accuracy of approximations. Moreover, an increase in hop number or $\mu$ results in a higher non-null BER level under \textbf{Case I} than for increases in $m$ under \textbf{Case II}.  \begin{figure}[!htb]
\centering\includegraphics[clip, trim=0.25 6cm 0.25cm 6cm,width=21pc]{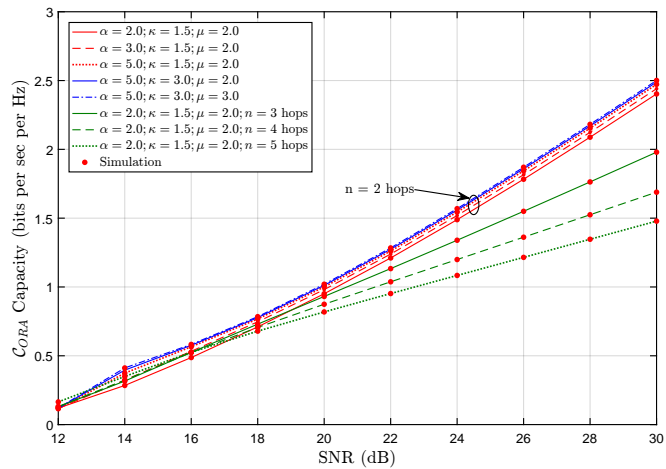}
  \caption{$\boldsymbol{\mathcal{C}}_{O}$ curves for the $\alpha-\kappa-\mu$ case showing the effect of increasing number relays and fading model parameter values under moderate fading conditions}\label{fig:12}
\end{figure} \begin{figure}[!htb]
\centering\includegraphics[clip, trim=0.25 6cm 0.25cm 6cm,width=21pc]{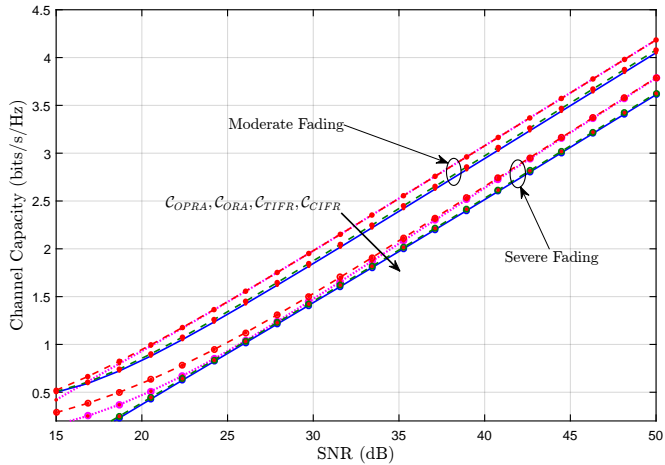}
  \caption{Channel capacity curves for the $\alpha-\kappa-\mu$ case under various adaptive transmission schemes for a 3-hop DF-based system}\label{fig:14}
\end{figure} Figure \ref{fig:12} presents the channel capacity under ORA adaptive transmission for \textbf{Case I}. There is improvement in channel capacity for increments in any of the fading model parameter values in Figure \ref{fig:12} in moderate fading. However, the improvements are very thin and even larger increments in parameter values do not cause a bigger gap between the curves. The gap between OPRA and ORA capacities gets smaller as SNR increases. Based on the analyses for AF, OP and BER, a similar behavior is expected for the other three transmission protocols. However, there appears to be an a diminishing reduction in channel capacity as more relays are added to the DF-based system. According to \cite{tha2017performance} this is due to the half-duplex capability of the relays which employs time-sharing multiple-access mechanism. Consequently, increasing the number of hops will require more time to deliver information from $S$ to $D,$ which hampers system capacity. Finally, Figure \ref{fig:14} shows the channel capacity under all the adaptive transmission schemes for moderate fading conditions. The results are in agreement with the findings of \cite{zhang2012performance} and prove that as expected $\boldsymbol{\mathcal{C}}_{P}\geq\boldsymbol{\mathcal{C}}_{O}\geq\boldsymbol{\mathcal{C}}_{T}\geq\boldsymbol{\mathcal{C}}_{c}.$ We expect similar results to the trend set in the preceding analyses for \textbf{Case II} and we therefore avoid displaying the figure for the sake of brevity.}
\section{Conclusions}\label{sec:6}
In this paper, various performance measures of a DF-based multi-hop system over generalized $\alpha-\kappa-\mu$ and $\alpha-\kappa-\mu$-Extreme fading channels have been examined in detail. We have provided the closed-form expressions for the OP, amount of fading, BER and channel capacity under various adaptive schemes for an end-to-end DF-based multi-hop relaying system. The effects of the variation of the fading parameters and the number of hops on the system performance is demonstrated.  System performance was shown to improve in a diminishing manner as the number of hop links increased under moderate fading. However, for severe fading conditions, there is little gain or actual performance degradation. From the analyses carried out in the present work, a few points which could also be of interest for further research are : (i) Determining the number of hops at which a negligible gain in system performance kicks in so as to avoid futile additional relaying if the purpose of relaying is to improve system performance. (ii) Determining the exact limit of BER under non-coherent modulation schemes as in \cite{rabelo2009mu} for the present fading models.
\bibliographystyle{IEEEtran}
\bibliography{IEEE_TVT_Paper11.bib}
\end{document}